\documentclass[twoside]{emulateapj}
\usepackage{apjfonts}
\usepackage{graphicx,natbib}
\usepackage{hyperref}
\bibpunct{(}{)}{;}{a}{}{,}
\let\tmpclearpage\clearpage
\let\clearpage\relax

\newcommand{\narrowfig}[3]{%
\begin{figure}[tbp]%
\begin{center}%
\includegraphics[width=86mm]{#1}%
\caption{#3}%
\label{#2}%
\end{center}%
\end{figure}%
}

\newcommand\FeI{\mbox{Fe\,\textsc{i}}}
\newcommand\NaID{\mbox{Na\,\textsc{i}\,D\ensuremath{_1}}}
\newcommand\mHz{\ensuremath{\mathrm{mHz}}}
\newcommand\km{\ensuremath{\mathrm{km}}}
\newcommand\mA{\ensuremath{\mathrm{m\AA}}}
\newcommand\nm{\ensuremath{\mathrm{nm}}}
\newcommand\s{\ensuremath{\mathrm{s}}}

\begin{document}

\title{On the propagation of $p$-modes into the solar chromosphere}
\shorttitle{$p$-mode propagation}
\author{A.G.~de~Wijn and S.W.~McIntosh}
\email{dwijn@ucar.edu}
\affil{High Altitude Observatory, National Center for Atmospheric Research\altaffilmark{1}, P.O. Box 3000, Boulder, CO 80307, USA}
\shortauthors{De Wijn et al.}
\and
\author{B. De Pontieu}
\affil{Lockheed Martin Solar and Astrophysics Laboratory, Palo Alto, CA 94304, USA}
\altaffiltext{1}{The National Center for Atmospheric Research is sponsored by the National Science Foundation.}

\begin{abstract}
We employ tomographic observations of a small region of plage to study the propagation of waves from the solar photosphere to the chromosphere using a Fourier phase-difference analysis.
Our results show the expected vertical propagation for waves with periods of 3 minutes.
Waves with 5-minute periods, i.e., above the acoustic cut-off period, are found to propagate only at the periphery of the plage, and only in the direction in which the field can be reasonably expected to expand.
We conclude that field inclination is critically important in the leakage of $p$-mode oscillations from the photosphere into the chromosphere.
\end{abstract}


\maketitle

\section{Introduction}\label{sec:introduction}

The chromosphere appears as a dynamic, thin, barbed red ring around the moon about the time of total solar eclipse.
Early detailed observations by, e.g.,
	\cite{Secchi1877},
have puzzled solar astronomers ever since.
We know now that the magnetized chromosphere is dominated by ``spicules'' at the limb and their on-disk counterparts
	\citep[``mottles'' and ``fibrils'',][]{1968SoPh....3..367B}.
Over 90\% of the non-radiative energy going into the outer atmosphere is deposited in the chromosphere
	\citep{1983ApJ...267..825W}.
The chromosphere requires nearly one hundred times the mass and energy flux of the corona for sustenance
	\citep{1977ARA&A..15..363W}
and remains the most poorly understood region of the outer atmosphere.
The complex dynamic appearance of the chromosphere, at the interface of photospheric and coronal plasmas where (on average) the magnetic and hydrodynamic forces balance, is incredibly difficult to interpret unambiguously and is largely responsible for the paucity of our knowledge.

Recent analyses of observations of the chromosphere at high spatial and temporal resolution made at multiple heights, guided by state-of-the-art numerical simulations, have begun to unlock the mysteries of the dynamic chromosphere
	\citep{2004Natur.430..536D,
	2006ApJ...647L..73H,
	2006ApJ...647L..77M,
	2006ApJ...648L.151J,
	2007ApJ...655..624D,
	2007ApJ...660L.169R,
	2007PASJ...59S.655D}.
Together, these results build a unified picture of the formation of (at least one type of) spicule resulting from the leakage of global $p$-mode oscillations along continuously evolving magnetic field.

In the traditional view of the solar atmosphere, 5-minute $p$-mode oscillations should not propagate upward because the acoustic cutoff is at periods of 3 minutes.
However, it has long been known that $p$-mode oscillations propagate upward in and around magnetic flux concentrations
	\citep[e.g.,][]{1978SoPh...58..347G}.
Several mechanisms have been proposed to explain this upward propagation of oscillations that have frequencies below the canonical values for the cut-off frequency in the photosphere, e.g., leakage along inclined field lines
	\citep[e.g.,][]{1977A&A....55..239B,
	1984A&A...132...45Z},
or radiative losses in the photosphere
	\citep{1983SoPh...87...77R}.
Because of a lack of high-resolution observations and modeling, this issue has not been resolved yet.

Recently, the advent of higher resolution observations and modeling have led to renewed interest in this topic, with suggestions that $p$-mode leakage can lead to formation of spicules
	\citep[e.g.,][]{1990LNP...367..211S,
	2004Natur.430..536D,
	2007ApJ...655..624D,
	2006ApJ...647L..73H,
	2007ApJ...660L.169R,
	2007ApJ...666.1277H}
and 5-minute ocillations in coronal loops
	\citep{2005ApJ...624L..61D},
and that it has important consequences for the energetics of the atmosphere
	\citep{2006ApJ...648L.151J}
and the damping of $p$-mode oscillations
	\citep{2007SoPh..246...53D}.
This has caused a renewed focus on understanding the causes for the propagation of $p$-mode oscillations into the atmosphere, with numerical models investigating both proposed mechanisms
	\citep[e.g.,][]{2004Natur.430..536D,
	2007ApJ...666.1277H,
	2008ApJ...676L..85K}.
While improvements in numerical models are important for a better understanding of this issue (especially with respect to the role of wave-mode coupling at the $\beta=1$ surface), the advent of space-based observations at higher spatial resolution allow for new and crucial observational constraints which can guide us towards a resolution of this issue.
In this Letter, we will provide observational evidence that $p$-mode oscillations propagate preferentially along inclined field at the periphery of plage.

\narrowfig{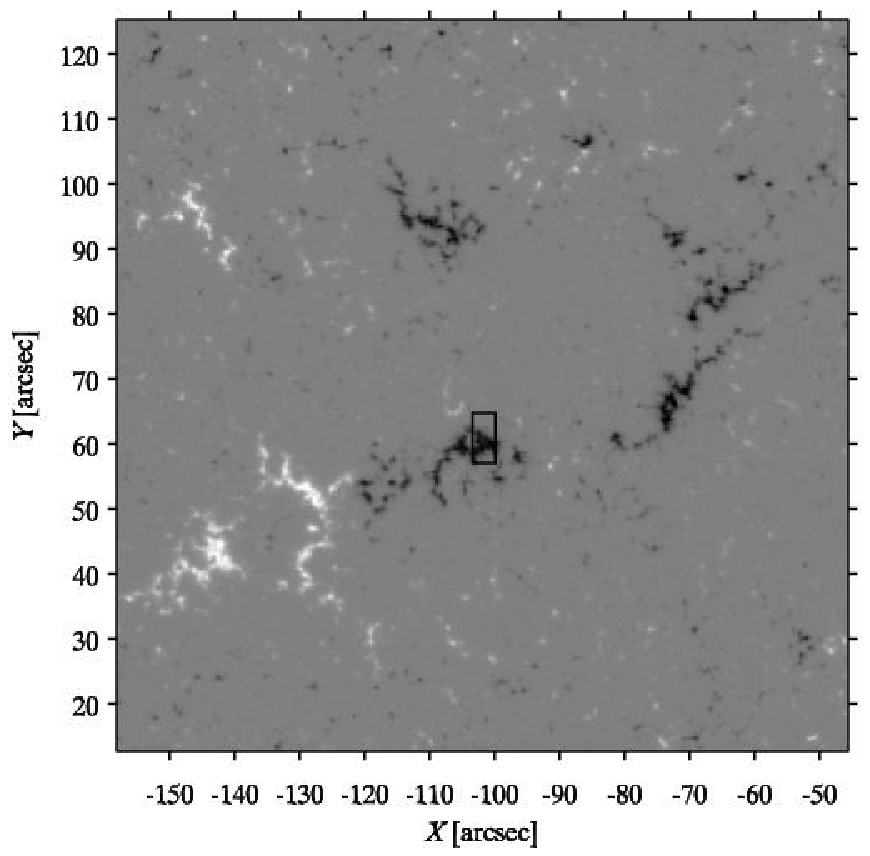}{fig:synop}{\NaID\ $V/I$ image of the region taken at 08:25:02~UT, several hours before the start of the sequence used in this Letter.
The approximate area shown in Figs.~\ref{fig:plage3} and~\ref{fig:plage5} is indicated by the black box.}

\section{Observations and reduction}\label{sec:observations}

We employ data sequences taken with the space-borne observatory \emph{Hinode}
	\citep{2007SoPh..243....3K}.
The Solar Optical Telescope
	\citep[SOT,][]{2008SoPh..249..167T,
	2008SoPh..249..197S,
	2008SoPh..249..233I,
	2008SoPh..249..221S}
was used to observe a small area of decaying plage close to disk center ($\mu=0.11$) on 2009 January~30 using both the Spectropolarimeter (SP) and the Narrowband Filter Imager (NFI).
A context image is shown in Fig.~\ref{fig:synop}.
SP was programmed to make repeating raster scans of $4.8\arcsec\times61.4\arcsec$, with a resolution of $0.16\arcsec$ and an integration time of $1.6~\s$ per slit.
NFI was concurrently used to observe 5 positions in the \NaID\ line, at $\pm168$, $\pm80$, and $0~\mA$ from the center of the line (see Fig.~\ref{fig:passband}).
The cadences of the SP rasters and NFI line scans are about $55~\s$ and $32~\s$, respectively.

\narrowfig{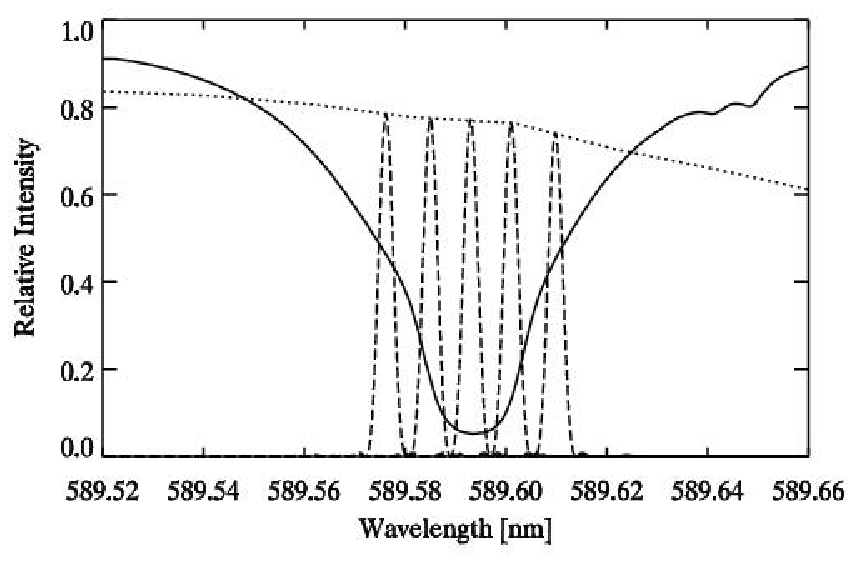}{fig:passband}{The \NaID\ line (solid line) together with filter transmission curves for a perfect Lyot filter using the NFI design at the 5 positions used in this analysis (dashed lines), scaled by the NFI \NaID\ prefilter transmission (dotted line).}

The data were carefully aligned using Fourier cross-correlation techniques.
First, the \NaID\ far and inner wings were aligned for each time step separately on the basis of their absolute Stokes $V$ signal, using the far blue wing image as reference.
Then, using the intensity image, the core of the line was aligned to the far blue wing.
The sequence was then aligned using the Stokes $V$ signal in the far blue wing.
The NFI wing data was then resampled using cubic spline interpolation to match the observing time of the core images.
Doppler velocities were calculated by fitting parabolas to each \NaID\ 5-point line scan.

For the SP data, a linear minus Gaussian function was fitted to each spectrum of the \FeI\ $630.25~\nm$ line.
The Doppler velocities derived from the repeating SP scans were interpolated in time to match the \NaID\ core images.
Finally, the SP data was mapped onto a grid matching the square NFI pixels of $0.16\arcsec$.
This entails scaling and warping of the SP data to correct for slit rotation and other distortions as well as the application of a fixed offset between the SP and NFI sequences plus the displacements computed for the \NaID\ core sequence in order to correct for image motion.
The fixed offset was computed by optimizing the correlation between the SP Stokes $V$ amplitude and the Stokes $V$ signal in the \NaID\ far blue wing.
The final sequence contains 211 frames.
A manual check shows an excellent alignment between the SP and NFI sequences.

\narrowfig{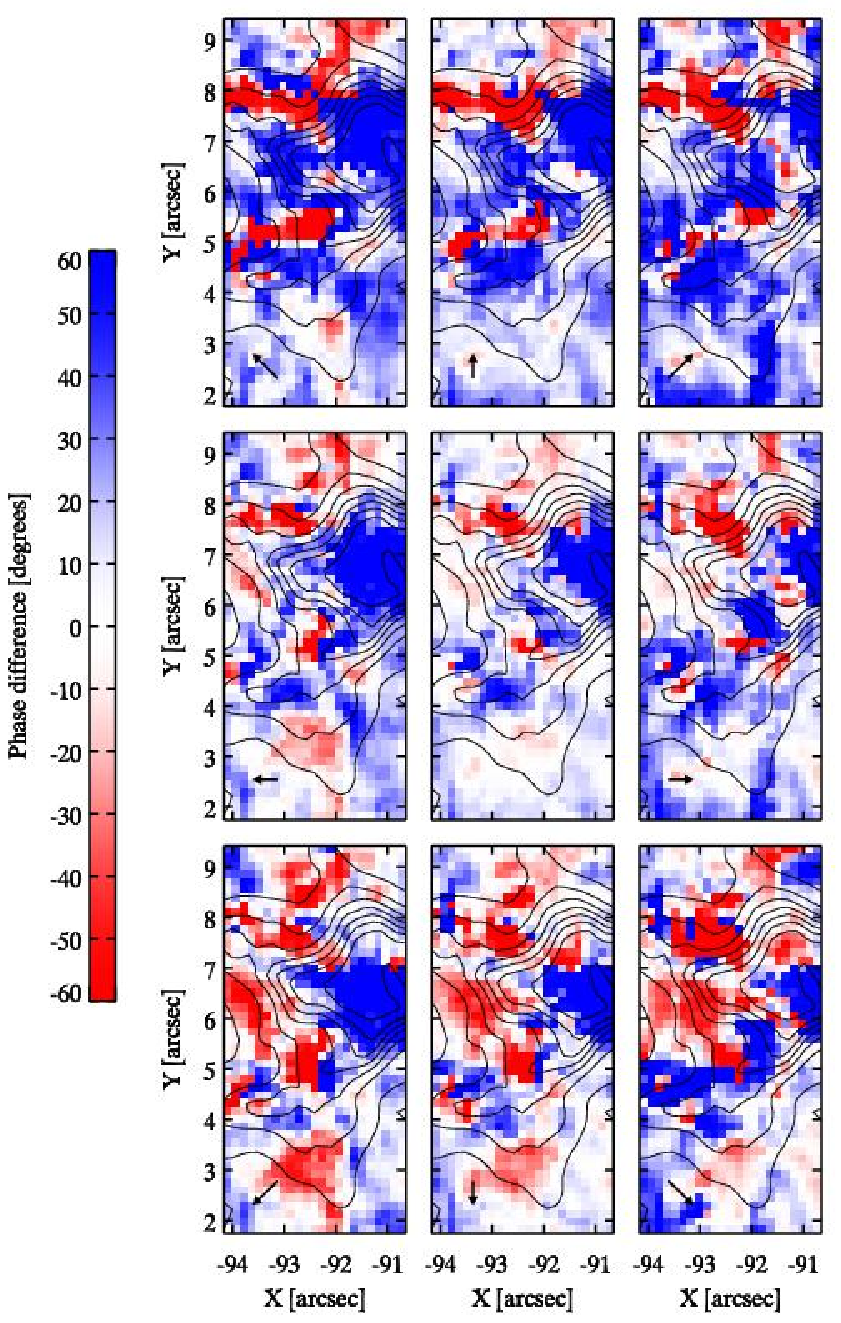}{fig:plage3}{Phase difference between Doppler shift in \FeI\ and \NaID\ at $5.6~\mHz$.
Blue and red respectively indicate positive and negative phase difference, i.e., upward and downward propagation.
Contours indicate the average Stokes $V$ signal in \FeI\ at $2.5\%$ intervals.
Arrows in the bottom left corner of each panel indicate the shift between \NaID\ and \FeI\ using \NaID\ as reference.
Rows are shifted in north by $3$, $0$, and $-3$ NFI pixels of $0.16\arcsec$, respectively from top to bottom.
Columns are shifted east by $3$, $0$, and $-3$ NFI pixels, respectively from left to right.}

\narrowfig{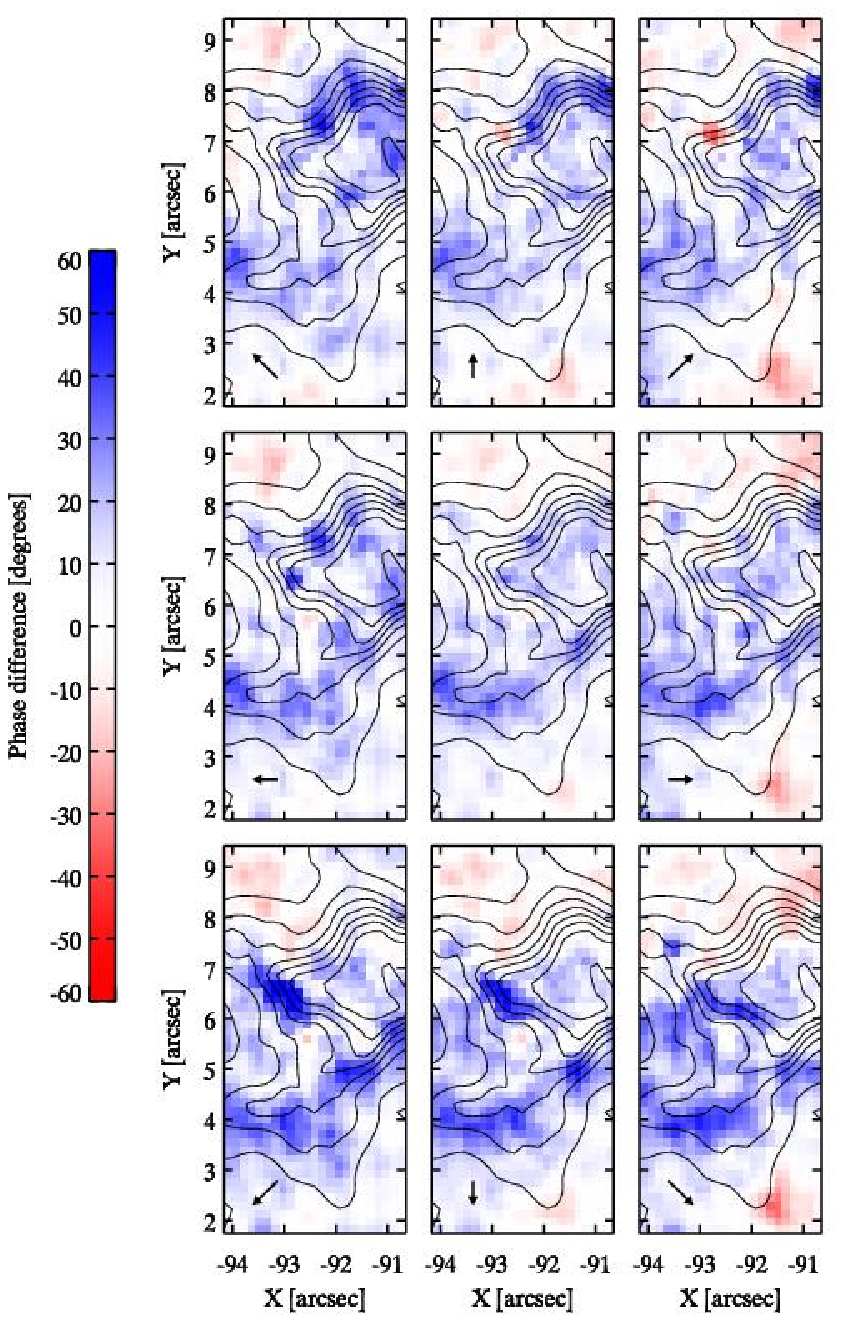}{fig:plage5}{Phase difference between Doppler shift in \FeI\ and \NaID\ at $3.3~\mHz$ in the same format as Fig.~\ref{fig:plage3}.}

We employ a Fourier analysis to derive phase differences between the Doppler velocities derived from the SP data and those from the NFI data.
The cadence of the SP observations limits the Nyquist frequency to $9~\mHz$.
A $\cos^2$ apodization window is applied to $10\%$ of the sequence.
In order to reduce noise, we average the phase difference by summation of crosspower weighted by a Gaussian with a FWHM of $1~\mHz$ centered on the frequency of interest.
Propagation along slanted rays is sampled by shifting the \NaID\ sequence by several pixels in each direction.

While the core of the \NaID\ line is expected to sample higher layers than the core of the \FeI\ line, it is not immediately obvious which Doppler velocity signal is formed higher.
In order to deduce empirically which is higher, we examine the phase difference of oscillations with $2\frac12$-minute periods.
Upward propagation is expected in the non-magnetic internetwork
	\citep[e.g.,][]{1993ApJ...414..345L}.
In the majority these areas, we find a positive phase difference (i.e., upward propagation) if we assume the \NaID\ Doppler velocity signal is formed lower than the \FeI\ signal.
We thus take the \NaID\ Doppler velocity signal as the reference.
The \FeI\ line is sampled with high spectral resolution, which allows for very accurate determination of the line core position.
Conversely, the relatively wide NFI passband, imperfect suppression of sidelobes, and the contribution of the \NaID\ wings in the determination of the Doppler velocity cause this diagnostic to sample deeper layers.

\section{Results and discussion}\label{sec:results}

\subsection{3-minute oscillations}

Figure~\ref{fig:plage3} shows the maps of phase difference at $5.6~\mHz$, corresponding to a period of 3~minutes.
The central panel shows propagation directly upward in the center of the plage region.
The general pattern remains the same under northward shifts, though upward propagation becomes more pronounced over a larger area.
There is less propagation in the core of the plage when sampling in south-west direction.
By introducing a sufficiently large shift, patches of downward propagation appear.

While the specifics of these results depend on the magnetic configuration of the plage, they are not unexpected.
Acoustic waves with 3-minute periods can propagate in vertical structures in the traditional treatment of wave propagation in the solar atmosphere.
Many previous studies of oscillations have shown that this indeed happens in both network and internetwork areas
	\citep[e.g.,][]{1995ESASP.376a.151R,2004A&A...416..333R}.
Oscillation power is reduced inside the network compared to the internetwork, but enhanced in the area immediately surrounding the network
	\citep[e.g.,][]{2001A&A...379.1052K}.
No obvious phase-difference signal that would correspond to these ``power aureoles'' is present.

We find much larger phase difference inside the plage region than outside it.
	\cite{2005A&A...430.1119D}
analyzed TRACE UV continua
	\citep{1999SoPh..187..229H},
and noticed a peak in their phase-difference $(k_h,f)$ diagram at $6~\mHz$ and at small spatial scales reaching 110~degrees, and also a corresponding peak in their spatially-averaged phase-difference spectrum of network areas at the same temporal frequency.
While those authors conclude that these features are likely an artifact in their data, the analysis presented here shows a conspicuously similar signature while using wholly different observations.
We thus conclude that the large phase difference measured inside the magnetic region here and also by
	\cite{2005A&A...430.1119D}
is not an artifact of the data or processing, but must be attributed to solar 3-minute waves in magnetic regions.
These waves must be compressible, since they would not show up as an intensity-modulation the TRACE UV continua otherwise.
Due to the large phase difference, slow mode magneto-acoustic waves are a good candidate.
Further study of 3-minute waves in regions of strong, vertical field through both observations and simulations is warranted.

\subsection{5-minute oscillations}

Figure~\ref{fig:plage5} shows the maps of phase difference averaged over a $1-\mHz$ range around $3.3~\mHz$, corresponding to a period of 5~minutes.
In the central panel, only very little propagation is detected.
None of the panels show significant propagation in the core of the plage region.
Introducing a shift invariably results in phase difference indicative of propagation in those areas of the plage region where the field is expected to diverge in the direction of the displacement.
As an example, the left-upper panel shows enhanced phase difference in the north-east corner of the plage region.
Conversely, the bottom row shows increased phase difference on the southern edge.
Waves with 5-minute periods propagate predominantly in those places where the field is inclined in the direction of propagation.

It is possible that the $\beta=1$ surface plays an important role in $p$-mode leakage into the chromosphere.
The $\beta=1$ surface is expected to intersect the photosphere at the periphery of plage.
If mode coupling is important in $p$-mode leakage into the chromosphere, one would thus expect it to happen preferentially at the edges of plage, consistent with the current results.

Only weak indications of downward propagation are detected, mostly at large shifts, e.g., at $(-92.5,7)$ in the upper-right panel.
Potentially, signals of downward propagation are not detected because they are overwhelmed by signals of upward propagating waves.
If a weaker signal of propagation in the opposite direction is present, the phase difference will be underestimated.
In the case where both signals are of equal amplitude, they can be considered as a standing wave with zero phase difference.

Because of the uncertainty in the sampling heights of the two diagnostics used, we should consider that in areas with strong magnetic field the heights could be reversed as compared to in internetwork.
However, in such a case, the phase difference would be negative, indicative of downward propagation, which conflicts with other studies
	\cite[e.g.,][]{2007ApJ...654.1128D}.
While we cannot deduce a propagation velocity from the phase differences measured, we are confident that we indeed detect upward propagation.

Studies by
	\cite{2006ApJ...640.1153C,
	2009ApJ...692.1211C}
show propagation in apparently vertical structures in a plage region.
Based on these observations,
	\cite{2008ApJ...676L..85K}
suggest that radiative relaxation is key to the propagation of 5-minute oscillations.
Our analysis is based on observations with much higher spatial resolution.
Conceivably,
	\cite{2006ApJ...640.1153C}
observed slanted propagation inside their resolution element, however, inversions of their spectropolarimetric measurements indicated that the field was truly vertical.
In the results presented here no propagation is found in areas where the field can be reasonably expected to be vertical, suggesting strongly that inclination is critically important for the propagation of 5-minute oscillations, at least in this plage region.

Given the results presented in this Letter, one may theorize that apparently vertical propagation is possible provided the field is sufficiently twisted.
The field is then inclined with respect to the gravity vector, and oscillations may travel upward along this field.
A structure with radius $r$ reaches an inclination of $\phi$ at its periphery if it makes at least one complete twist over a height of $h=2\pi r\tan(\phi)$.
Since we do not accurately know the difference in the formation height of the observations used here, we cannot estimate the angle of propagation from these data.
However, in order to adequately lower the acoustic cut-off, $\phi\ge30^\circ$ is needed
	\citep{2004Natur.430..536D}.
Assuming a constant $r=50~\km$, we find $h\le180~\km$.
As the fluxtube expands with height, twist and hence inclination are enhanced naturally, so that $h$ may be significantly increased
	\citep[e.g.,][]{1974ApJ...191..245P}.
Measuring twist in such a small structure is difficult, and so it is not immediately clear that this limiting value for $h$ is reasonable.
However, since we do not see significant propagation inside the plage region, it seems that at least in this case, there is insufficient twist to allow for apparently vertical propagation of $p$-mode oscillations.

\section{Conclusions}\label{sec:conclusions}

We have studied the propagation of waves in a small region of plage using a Fourier phase-difference analysis.
We find expected behavior of 3-minute oscillations.
These oscillations are seen to propagate vertically, as is expected in the traditional view of the propagation of acoustic waves in the solar atmosphere.
We find a much larger phase difference for these oscillations inside the plage region than outside it.
Our results show that propagation of 5-minute $p$-mode oscillations happens only along inclined field at the periphery of the plage region.
We find no propagation of $p$-modes in the core of the plage, where the field is expected to be mostly vertical.

\acknowledgments{We are grateful to B.~W.~Lites for discussions and his assistance in reducing the \emph{Hinode}/SP data.
BDP is supported by NASA grants NNM07AA01C (Hinode), NNG06GG79G and NNX08AH45G.
\emph{Hinode} is a Japanese mission developed and launched by ISAS/JAXA, with NAOJ as domestic partner and NASA and STFC (UK) as international partners.
It is operated by these agencies in co-operation with ESA and NSC (Norway).}

\let\clearpage\tmpclearpage

\end{document}